\documentstyle[12pt]{article}
\def\Cal#1{{\cal#1}}
\def\BE{\begin{equation}}
\def\EE{\end{equation}}
\def\Dx{{\partial}_x}
\def\Dy{{\partial}_y}
\title{FORCES ON BINS:   \\
 THE EFFECT OF RANDOM FRICTION\footnote{Research supported in part by the
National Science Foundation under grant DMS9504433.}}
\author{ E. Bruce Pitman  \\
Department of Mathematics\\
State University of New York\\
Buffalo, NY, 14214\\
pitman@galileo.math.buffalo.edu
}
\date{Revised: 18 November, 1997}
\begin{document}
\maketitle

%
%
%
%
%
%
%
%




\noindent
{\bf ABSTRACT}.  The q-model of Coppersmith et al.~has 
renewed interest in understanding the
forces generated along the walls and at the bottom of a silo filled with a 
granular material.  Fluctuations in the mean stress have been characterized
for the q-model, and related to experimental work on stress chains.
The classical engineering approach to bin loads follows from
Janssen's analysis, predicting a saturation of stress, as a
function of depth, in a tall silo.
In this note we re-examine the Janssen theory, introducing
randomness into the important parameters in the theory.
The Janssen analysis relies on assumptions not met in practice.  For this
reason, we numerically solve the PDEs governing
the equilibrium of forces in a bin, 
again including randomness in parameters.
We show that 
the most important of these parameters is a coefficient of friction at the
wall of the bin.
This random friction model combines some features of fluctuations 
as seen in experiments, with a classical continuum mechanics approach to 
describing granular materials.
\vspace{3.0in}
\eject

\noindent
{\bf 1. INTRODUCTION} The classical engineering theory of Janssen
\cite{jansen}
provides an estimate for the mean vertical stress in a silo filled with a 
granular material.  The principal feature of the Janssen analysis is that,
under passive stress conditions, the
mean stress saturates, asymptoting to a value depending on bin radius
and wall
and internal friction coefficients, but independent of the height.
The Janssen theory relies on two assumptions, assumptions
which do not hold in practice.  Nevertheless, the analysis gives a
reasonable estimate of the bin loads, and its simplicity is its virtue.
Several analyses have attempted to remove some of
the assumptions of the Janssen
theory; the interested reader should consult \cite{neddermanbook}.

Recently Coppersmith et.al. \cite{sue1,sue2} have developed a
model for the force distribution in a bin.  In this ``q-model'',
any one particle within the sample transmits its weight to neighbors
that are below, in a random manner. The authors 
derive a mean field theory based on this model, and
find fluctuations in the forces 
felt by the lowest row of particles.  Under most assumptions on
the choice of random number distribution,
the number of occurrences of a fluctuation of
a given size decays exponentially with size.

The current incarnation of the q-model is scalar: only the vertical
force is balanced.  
Recently, Socolar has introduced a generalization,
the so-called $\alpha$-model, which balances vertical and horizontal forces
and angular moment.  The $\alpha$-model contains three random variables,
and analysis appears difficult.  However numerical simulations
modeling rough walled bins appear consistent with the classical
continuum theory; numerical simulations modeling infinitely wide bins
appear consistent with the q-model for a special distribution of the q's.

To provide a framework for introducing fluctuations into
a continuum setting,
we incorporate some of the randomness of the q-model into the
Janssen analysis.  In Section 2 we reconsider the Janssen derivation
and include a random component into
the grain friction, and reformulate the balance law as a stochastic
differential equation.  Standard results of stochastic calculus
provide an estimate of the mean stress and its
variance, at any height.
In Section 3, we numerically solve the complete stress equilibrium equations,
assuming a Mohr-Coulomb constitutive relation, and
again including a random component in the friction.  
Under passive loading, the stress saturates; stress fluctuations 
are not significant until near saturation. 

An experimental
finding closely related to the current note is \cite{clementpng}.  
That paper reports careful measurements of force
fluctuations in tall narrow bins,
bins whose widths range from three to eight grain
diameters and whose depth ranges up to about 100 grains diameters.
Measured average vertical stress at any depth is
systematically higher than predicted by
the Jansen theory, and
fluctuations in this stress range up to about 20\%.  These
fluctuations are apparent only after the stress starts to saturate.
(N.B. Socolar \cite{socolar} also finds the Janssen stress
smaller than his calculated average stress, at any depth.)
These experiments also demonstrated
a dependence of stress on ambient temperature, an effect
we do not consider here.
\vspace{0.25in}

\noindent
{\bf 2. GENERALIZED JANSEN ANALYSIS} We briefly review Janssen's
theory, and provide a stochastic generalization of that analysis.
See \cite{neddermanbook} for the fundamental mechanics of granular media.
All of this study is restricted to two space dimensions.

Let the average vertical stress be
denoted $ \bar \sigma ~=~ \int_{-D/2}^{D/2} \, \sigma^{yy}(x,y) \, dx $,
where $\sigma^{xx}, \sigma^{xy},\sigma^{yy}$ are the
$xx$, $xy$, $yy$-components, respectively, of the (symmetric) stress tensor
$T$.

\begin{figure}
\vspace{3.0in}

\includegraphics{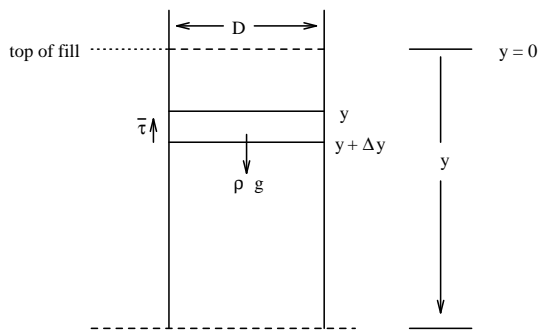}
\caption{Force balance for Janssen's analysis.  On the slice,
stresses and gravity are balanced by wall friction.}
\end{figure}

Consider the force diagram in Figure 1; at
equilibrium, the average stress at
$y$ and $y+\Delta y$, gravity, and wall friction $\bar \tau$ are balanced:
\begin{equation}
\Dy \bar \sigma ~+~ \frac{2\bar \tau}{D} ~=~ \rho g  ~~.
\end{equation}
Now we make two assumptions, critical to the Janssen theory, but which
do not hold in practice:
\begin{enumerate}
\item{At every point $\sigma^{xx}$ and $\sigma^{yy}$ are the principal
stresses (i.e., the eigenvalues of the stress tensor)
and the Coulomb frictional condition implies
$ \sigma^{xx}(x,y)=K \sigma^{yy}(x,y)$,
$K= \frac{1+s}{1-s}$, $s=\sin(\phi)$ and
$\phi$ is the internal friction
angle;}
\item{Along the wall, $\bar \tau = \sigma^{xy} (\pm D/2,y) = \delta
\sigma^{xx} (\pm D/2,y)$ where $\delta =
\tan(\phi_w)$, $\phi_w$ is the wall-material friction angle}
\end{enumerate}
\noindent
Combining these assumptions, we arrive at the equation
\begin{equation}
\Dy \bar \sigma ~+~ \alpha \sigma ~=~ \rho g,~~~~~~~~
\alpha= \frac{2\delta K }{D}~~ .
\label{janseneqn}
\end{equation}
Solving subject to $\bar \sigma \rightarrow 0$ as
$y \rightarrow 0$ gives
\begin{equation}
\bar \sigma (y)  ~=~ \frac{\rho g }{\alpha}
\left(
1-\exp(-\alpha y)
\right) ~.
\label{jansensoln}
\end{equation}
It is apparent that the average stress saturates, the
asymptotic value $\frac{\rho g }{ \alpha}$ depending on the
material and wall parameters and the bin diameter.

The formula for $K$ is based on the assumption that the stress field
is in the passive state, with the $xx$-stress the major principal stress
(the larger of the eigenvalues) and the $yy$-stress the minor (the smaller
eigenvalue).  If the material is in the active state, the
$yy$-stress is major, the $xx$-stress minor, and $K$ is replaced by
$K^{-1}$.
For a typical material, $\phi$ may be 30$^{\circ}$, so $K = 3$ in the
passive state.  In the active state this parameter is $\frac{1}{3}$,
and saturation of the stress requires a bin that is an order of magnitude
taller.

Now assume the coefficient of the stress in (\ref{janseneqn})
has a mean and
a fluctuating component.  
This fluctuating component might arise from randomness in the
friction angle, for example.  Assuming an It$\hat o$ formulation for
the resulting stochastic differential equation, write
\begin{equation}
d \bar \sigma ~=~- \alpha \bar \sigma dy 
- \epsilon \bar \sigma dW + \rho g dy ~.
\label{randomjansen}
\end{equation}
Here $dW(y)$ is a Wiener measure associated with the random fluctuations,
and $\epsilon$ is a measure of the size of the fluctuations.
Standard arguments give the following results (see, e.g., chapt. 8 in
\cite{arnoldbook}).
A formal solution may be obtained by a variation of parameters argument,
but more insightful are formulae for the first and second
moments.  The mean of the solution,
$m \doteq \Cal E ({\bar \sigma})$,
is, not surprisingly, the Janssen solution (\ref{jansensoln}).  The
second moment
$P \doteq \Cal E ({\bar \sigma}^2) $, 
satisfies 
$$
\dot P = (-2 \alpha + {\epsilon}^2) P
+ 2 m \rho g ~.
$$
Thus 
\BE
P = \frac{2 m \rho g}{2 \alpha - {\epsilon}^2}
[1 - \exp (-(2 \alpha - {\epsilon}^2) y)]
\EE
The standard deviation is $\sqrt{P-m^2}$, and 
an order of magnitude estimate gives the deviation $\sim
\frac{m \epsilon}{\sqrt{2 \alpha}}$, after the stress has saturated.

An alternative hypothesis is that 
randomness in packing leads to fluctuations in the density, and
thus to fluctuations in the stress.  That is,
the weight $\rho g$ must include a random component due to voids.
This assumption leads to the equation
\begin{equation}
d \bar \sigma ~=~- \alpha \bar \sigma dy 
+ \rho g dy + \epsilon \rho g dW  ~.
\label{randomdensity}
\end{equation}
The mean of the solution is, again, given by (\ref{jansensoln}).
The standard deviation is 
$\frac{(\epsilon \rho g )}{\sqrt{2 \alpha}}
[1 - \exp (-2 \alpha y)]^{\frac{1}{2}}$.  
\vspace{0.25in}

\noindent
{\bf 3. EQUILIBRIUM ANALYSIS } The Janssen analysis relies on 
assumptions not met in practice.  In this section we solve the full
stress equilibrium equations for a Coulomb material in a bin.  Although
analysis is possible in the limiting case of smooth walls
(see \cite{neddermanbook}), this section determines solutions numerically.

Stress equilibrium is written
\begin{eqnarray}
\Dx \sigma_{xx} + \Dy \sigma_{xy} &=& 0  \\
\Dx \sigma_{yx} + \Dy \sigma_{yy} &=&  \rho g ~.
\label{stresequ}
\end{eqnarray}
A common constitutive assumption is that the material is
Mohr-Coulomb, at incipient yield.  That is,  
one assumes the ratio of the shear stress, $\tau$, to the mean stress,
$\sigma$, is a constant, where
\BE
\sigma = \frac{\sigma_1 + \sigma_2}{2}~~~~~~
\tau =   \frac{\sigma_1 - \sigma_2}{2}
\EE
\noindent
and $\sigma_1,~\sigma_2$ are the eigenvalues of the stress
tensor $T$.
The Mohr-Coulomb condition reads
\BE
\frac{\tau}{\sigma} = s ~.
\EE

The Mohr-Coulomb condition can be viewed as a nonlinear relation
for, say, $\sigma_{yy}$ in terms of $\sigma_{xx}$ and $\sigma_{xy}$.
It is often convenient to make a change of variables that incorporates
this relation.  With the mean stress $\sigma$ defined above, introduce
the angle $\psi$, measured from the horizontal,
such that $(\cos(\psi),~\sin(\psi))$ is an
eigenvector of $T$ associated with $\sigma_1$.
Then write
$$
T = \sigma
\left (
\begin{array}{cc}
1 & 0 \\
0 & 1
\end{array}
\right ) + \sigma s
\left (
\begin{array}{cc}
\cos(2 \psi ) &\,\,\,\,\,\, \sin(2 \psi ) \\
\sin(2 \psi ) & -\cos(2 \psi )
\end{array}
\right ).
$$
\noindent
This equation specifies the stresses in terms of two dependent variables,
$\sigma , \,\psi$, whose evolution is determined by the equilibrium
equations.

This change of variables may be used to rewrite the momentum equations as
\begin{eqnarray*}
\left (
\begin{array}{cc}
1+ s \cos(2 \psi ) &\,\,\,\, -2 \sigma s \sin(2 \psi ) \\
s \sin(2 \psi )  &~~ 2 \sigma s \cos(2 \psi )
\end{array}
\right )
&
\Dx
&
\left (
\begin{array}{c}
\sigma  \\
\psi
\end{array}
\right ) 
+                                     \\
\left (
\begin{array}{cc}
 s \sin(2 \psi ) &\,\,\,\, 2 \sigma s \cos(2 \psi ) \\
1 - s \cos(2 \psi )  &\, 2 \sigma s \sin(2 \psi )
\end{array}
\right )
&
\Dy
&
\left (
\begin{array}{c}
\sigma  \\
\psi
\end{array}
\right ) 
~=~ \left (
\begin{array}{c}
0  \\
\rho g
\end{array}
\right ) 
\end{eqnarray*}
\noindent
We nondimensionalize by scaling
length by the bin diameter $D$, and stress by $\rho g D$.  All calculations
are reported in non-dimensional units.
The independent variable
are $-1/2 \leq x \leq 1/2$ and $0 \leq y \leq H$.
This system of PDEs is strictly hyperbolic,
with characteristics inclined at an angle $\pm(\frac{\pi}{4}
- \frac{\phi}{2})$ from the direction of major principle stress.
The $y$-direction may be taken as the time-like direction.  ``Initial''
conditions for $\sigma$ and $\psi$
are imposed at the top of the fill, $y=0$, and the equations solved
downward.  At the boundaries, the bin walls
$x = \pm 1/2$, the wall friction angle
is imposed: $\psi = \delta$.
The system of equations is solved by a modification of the
TVD/Central Difference scheme of Nessyahu and Tadmor \cite{tadmor}.
The method is second-order accurate, and designed to avoid
spurious oscillations common to many higher-order schemes
for hyperbolic systems.
For the computations reported here, a gridsize of $\Delta x = 0.02$ was used.
On very coarse grids, fluctuations are larger than shown; after
sufficient refinement, the size of fluctuations appears to stabilize.

To introduce fluctuations, at each gridpoint at each level, the friction
angle is chosen with a random component.  Specifically, if $\phi$ is the
nominal friction angle, the angle used is
$\phi_{fluct}=\phi(1.0 + \zeta \xi)$ where $\xi$ is chosen randomly from a
uniform distribution between [-0.5,0.5] and $\zeta$ is an adjustable
parameter measuring the extent of variation in the friction angle.
Depending on testing apparatus, 
variations in measurements of the internal friction angle are as large as
$\pm 5^{\circ}$, more than $10\%$ of typical values
\cite{schwedes};
without good measurements of
the span of friction angles, we conservatively set
$\zeta = 0.1$.
So, for a nominal friction angle of 30$^\circ$, $\phi_{fluct} \in
[28.5,\,31.5]$; the sine of this angle is used in the constitutive
relation, and this sine $\in [.477,\,.522]$, about a $\pm 5\%$ swing.
Of course variations of friction in a real sample may have spatial
correlations; absent good modeling justification for a particular choice
of correlation, none is used here.
A random component of the wall friction angle (the boundary condition
$\delta$) is added in a manner similar to $\phi$.
We emphasize that our choice of $\zeta$ sets the imposed variation
in friction angle, and thus of the stress, but this choice is rather arbitrary.

The first result to understand is a typical stress profile, 
without any friction fluctuation, and the same parameters but
with fluctuation.  This is shown in Figure 2, which displays
the $yy$-component of the stress at the centerline of the bin and 
at the bin wall.  For comparison, the Janssen stress is also shown.
\begin{figure}
\vspace{3.95in}

\includegraphics{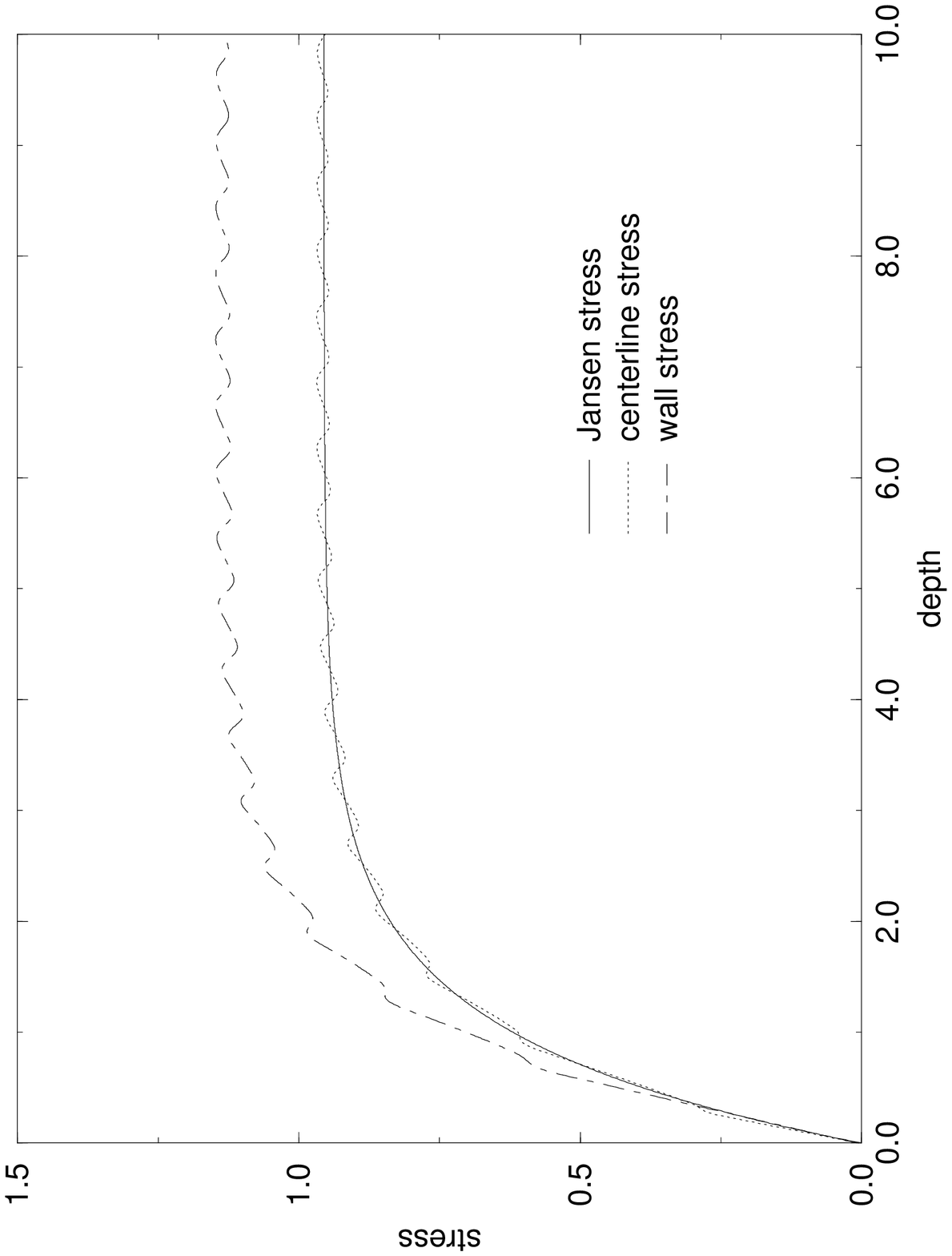}
\includegraphics{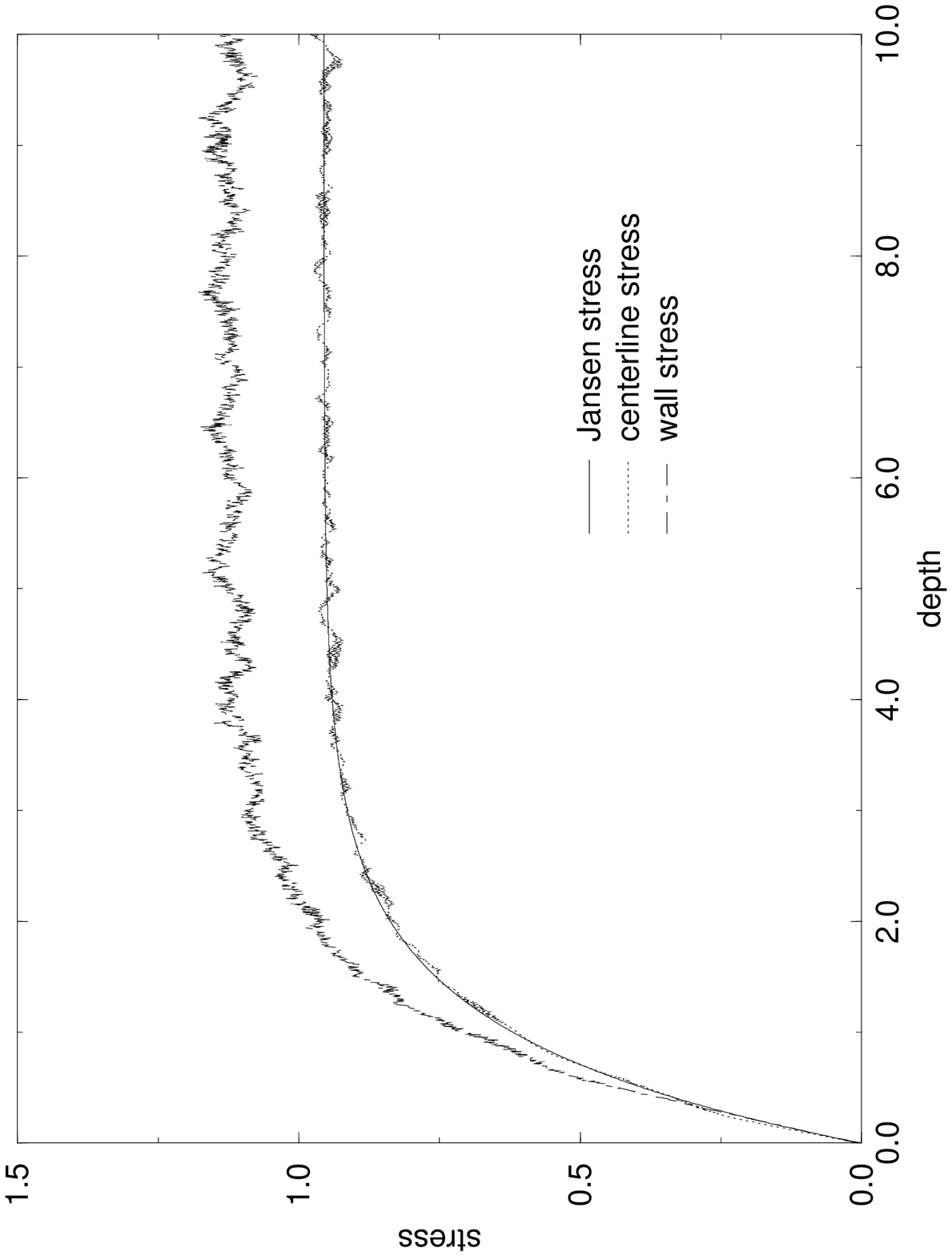}

\caption{(a) The $yy$-component of the stress at the
centerline and the wall, with no random
component of friction.  For comparison the Janssen solution is also
plotted.  Here the nominal internal friction angle $\phi = 30^{\circ}$ and the
nominal wall friction angle $\delta = 15^{\circ}$.
(b) Similar plot, but with a random component added to the friction angles.
}
\end{figure}
We have imposed the ``initial condition''
$\sigma = 0$ on $y=0$.  However, the condition for a surface
$y=h(x)$ to be stress free is (in general) inconsistent with
$y=const$; imposing $\sigma = 0$ leads to a free boundary problem
for the upper surface, a problem we do not wish to address here.
The regular oscillations in Figure 2(a) are due to mismatch in the
imposed stress at the
intersection of the $y=0$ surface and the bin wall, and are
well documented (see e.g. references in \cite{neddermanbook});
the period of these oscillations is related 
to the speed of the characteristics of the hyperbolic system.  In
Figure 2(b), fluctuations
at the walls are larger than at the centerline, and the wall stress
is some $15\%$ larger than centerline.
Notice that the regular
oscillations in Fig. 2(a) are dissipated by the randomness.

In Figure 3, the $yy$-stress is shown as a function of position across the bin,
at the depth $y=10$, the terminus of the computations in Figure 2.
\begin{figure}
\vspace{3.95in}

\includegraphics{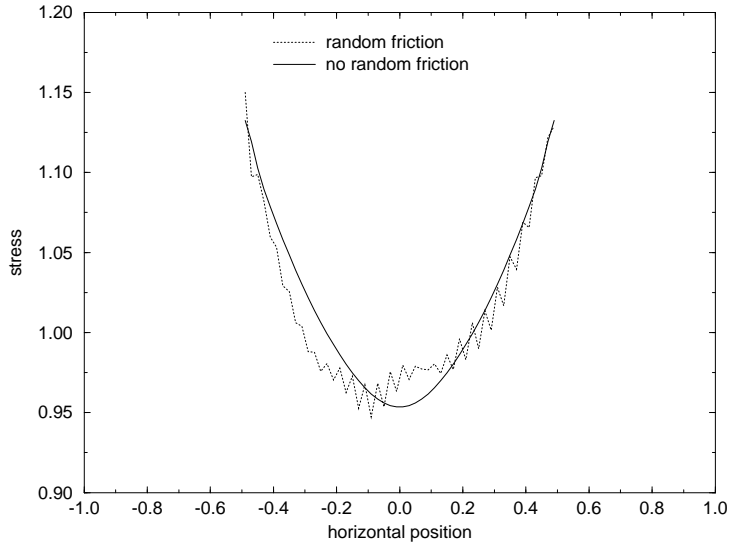}

\caption{Variation in the $yy$-stress across the width of
the bin, at $y=10$.  Shown are results both with and without
a random component of the friction angles.
In both cases, the nominal friction angles are
$\phi = 30^{\circ},~\delta=15^{\circ}$.
}
\end{figure}
The variation across the bin illustrates the limitations of the Janssen
assumptions.  Nonetheless, Figures 2 and 3 show that the Janssen analysis
provides a good estimate of the centerline stress (and NOT of the
average stress!).  This partially explains why
the measurements of \cite{clementpng} are larger than the Janssen predictions.
The centerline stress
is typically $15-20\%$ smaller than the largest stresses, found at the wall.

Figure 4 illustrates the sensitivity of computations to changes in the 
nominal friction angles.
\begin{figure}
\vspace{3.95in}

\includegraphics{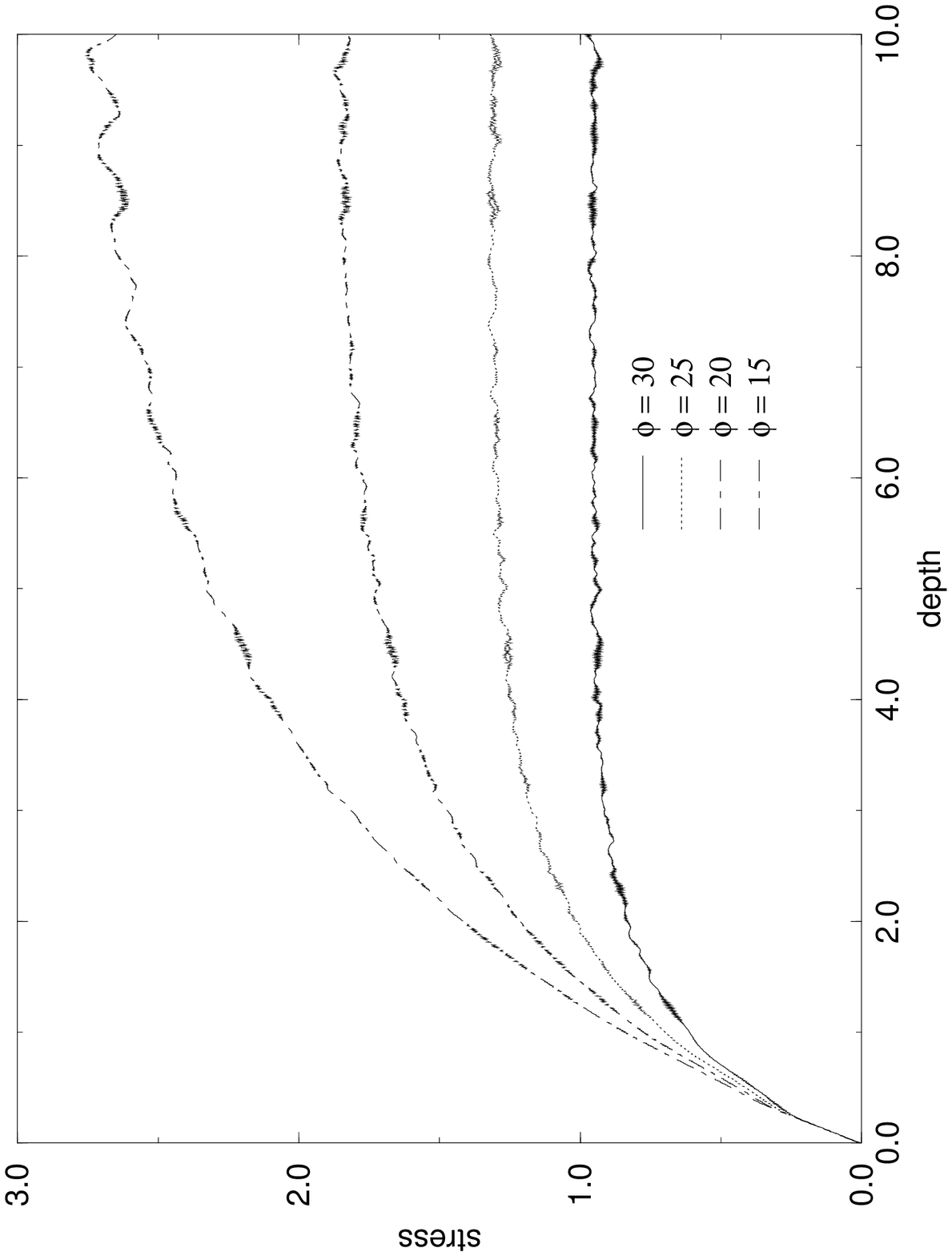}
\includegraphics{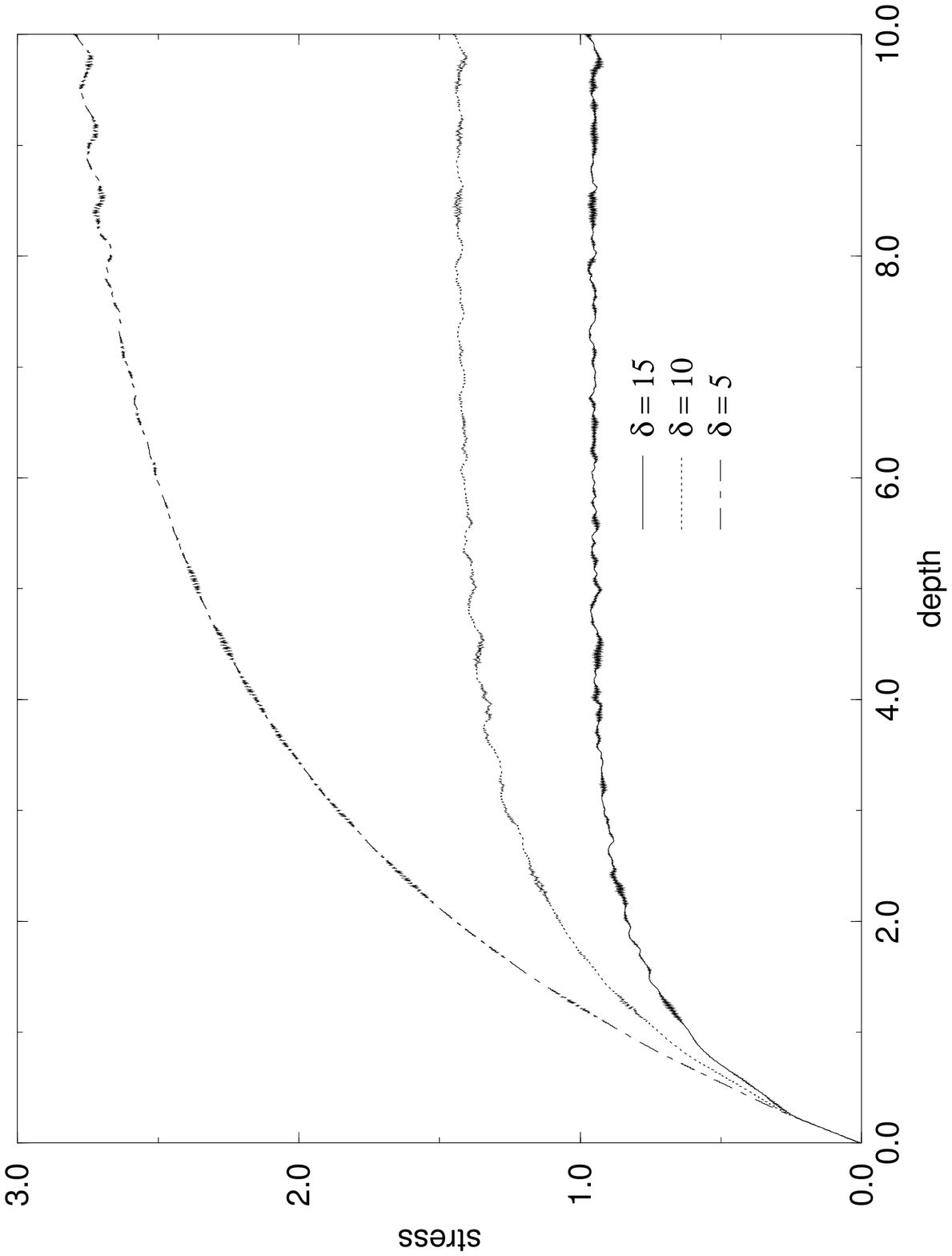}

\caption{(a) The $yy$-component of the stress at the
centerline for different nominal internal friction angles.  Nominal
wall friction is held fixed $\delta = 15^{\circ}$.
(b) The $yy$-component of the stress at the
centerline for different nominal wall friction angles, with fixed
nominal internal friction angle $\phi = 30^{\circ}$.
}
\end{figure}
In Fig. 4 (a), wall friction is held fixed while
the nominal internal friction angle $\phi$ is varied
from $15^{\circ}$ to $30^{\circ}$ (N.B. Recall that
the random fluctuation is $5\%$ of the nominal angle). 
With lower internal friction, fluctuations become more pronounced.  We 
conjecture that this is due to a
lower friction angle transmitting a smaller fraction of stress (and
of stress fluctuations) to
the walls, leaving a larger fraction of stress (and of stress
fluctuations) to be transmitted vertically.
Notice too that, at the smallest friction angle, the regular oscillations
of the stress reappear.
When internal friction is held constant but wall friction is varied,
Fig. 4 (b), the stress saturates deeper in the bin, and fluctuations
are not apparent until after this saturation.
We note that, with no 
wall friction, no weight is transferred to the bin walls and
a hydrostatic stress results.
Similarly, when periodic boundary conditions are imposed, 
a hydrostatic stress results.

In Figure 5, fluctuations for two sets of friction angles are
plotted.  In each case, the equations were solved through $y=50$.  The
centerline stress for $20<y<50$ was extracted, and the average computed;
this average should be the asymptotic value of the stress.  A normalized
deviation from the mean was found by
subtracting the mean from the sample value, and dividing by the mean.
For viewing, one signal is offset by 0.05.
\begin{figure}
\vspace{3.95in}

\includegraphics{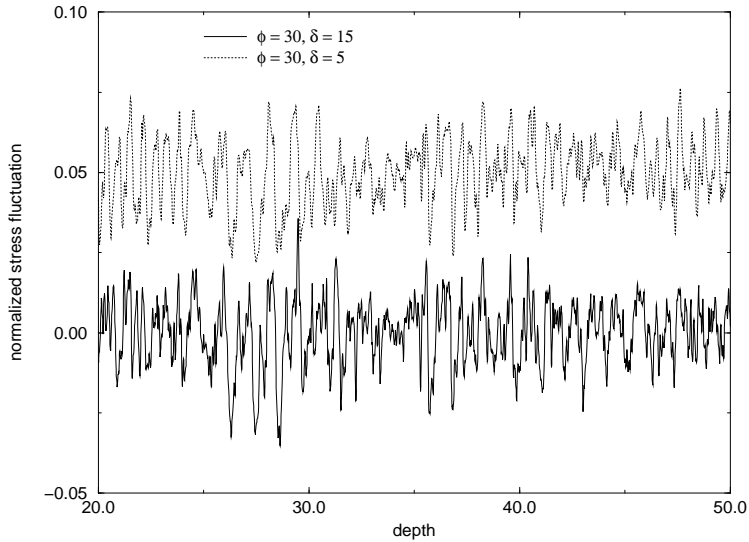}

\caption{Normalized fluctuations in the 
centerline $yy$-stress for two pair of friction angles.  Both signals are
demeaned; for viewing, the top signal is vertically offset by 0.05.
}
\end{figure}
For the baseline case $\phi=30^{\circ},~\delta=15^{\circ}$,
the internal friction angle varies by about $\pm 1.5^{\circ}$ and 
the wall friction angle by about $\pm 0.75^{\circ}$; the stress exhibits
fluctuations of about $\pm 4\%$.
For the case $\phi=30^{\circ},~\delta=5^{\circ}$,
the internal friction angle again varies by about $\pm 1.5^{\circ}$ but
the wall friction varies by only $\pm 0.25^{\circ}$; the stress 
fluctuates about $\pm 2.5\%$.

Figure 6 provides a plot of spectral power for the baseline case
$\phi=30^{\circ},\, \delta = 15^{\circ}$. 
The stress
was computed to a depth of $y=50$;
recall from Figure 2 that, for the
given friction angles, the stress
saturates well before $y=10$.
The centerline stress is sampled at every second timestep, from
about $y=20$ to $y=50$.
The power is computed using a Welch window with overlap, on the last
2560 sampled values.
\begin{figure}
\vspace{3.95in}

\includegraphics{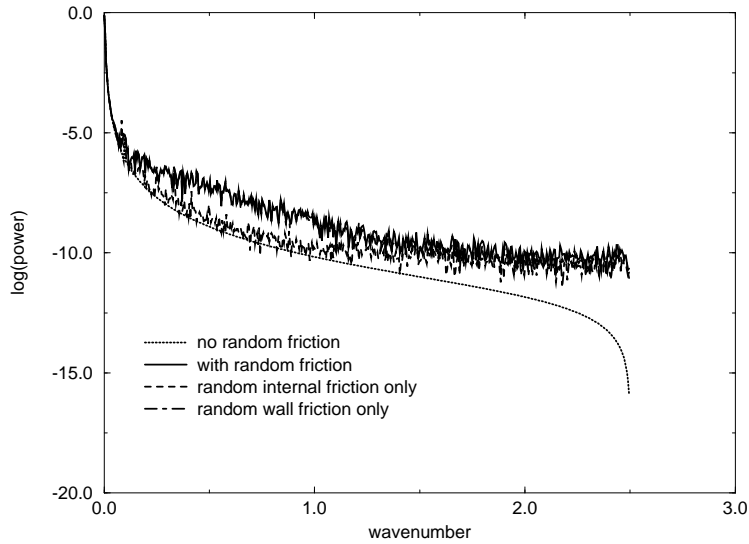}

\caption{Log (base 10) of the spectral power for four variations
of the base case.  The variations are
no random component of friction, a random component 
of both internal and wall friction,
a random component added to internal friction only, and
a random component added to wall friction only.
The nominal $\phi = 30^{\circ}, \, \delta = 15^{\circ}$. 
}
\end{figure}
Shown is the (base 10 log of the) power for four
variations: $(i)$ no random component of friction; $(ii)$ a random 
component of both internal and wall
friction; $(iii)$ a random component added to internal
friction only;
$(iv)$ a random component added to wall friction only.  The 
power for wall friction only lies atop the spectrum for wall and internal
friction.  The power for internal friction only deviates from these at
lower wavenumber.  Thus
fluctuations in the stress are essentially
due to a random component in the wall friction angle.  
From Figure 5, these fluctuations range up
to about $\pm 4\%$ of the mean.  Recall, this variation is based
on about a $5\%$ variation in the friction coefficient.  The
fluctuations reported in \cite{clementpng} are as large
as 20$\%$.  This comparison suggests that a
$15 - 20\%$ variation in the friction coefficient is not an unreasonable
parameter in stochastic models like the present.

Analysis of the q-model shows that 
the number of occurrences of a fluctuation of
a given size decays exponentially with size.
Recent experiments \cite{mueth} on short bins verify this finding,
for stresses larger than the mean; stresses smaller than the mean 
decay like a power law. 
\begin{figure}
\vspace{3.95in}

\includegraphics{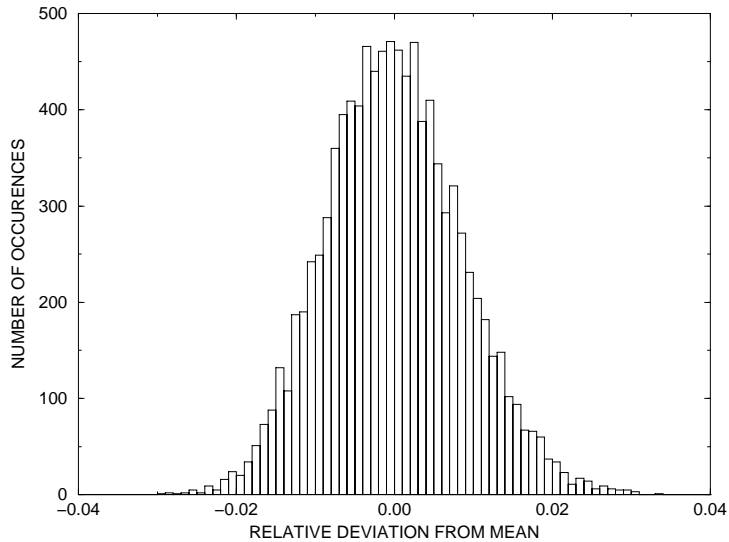}

\caption{A histogram of the normalized deviation of the
centerline $yy$-stress from the mean, evaluated at $y=50$.
Friction parameters were $\phi=30^{\circ},~\delta=15^{\circ}$.
Computed for 10,000 realizations, the distribution appears Gaussian,
not exponential as predicted by the q-model.
}
\end{figure}
Figure 7 presents the distribution for
the random friction model.  The equilibrium equations were solved 
to a depth
$y=50$, and the centerline stress was recorded.  Ten thousand 
realizations were made.  The average over
all realizations was calculated, subtracted from the sample value,
and this difference was normalized by the average.
Figure 7 is a histogram of these relative deviations.  The distribution of
fluctuations appears Gaussian, not exponential.
\vspace{0.25in}

\noindent
{\bf 4. Summary } We have reexamined the Janssen analysis
incorporating a random component of friction, solving for the
mean and the second moment of the stress.  For comparison, the nonlinear
equilibrium equations for a Mohr-Coulomb material
with random friction are solved numerically.  The analysis suggests that
fluctuations are
significant only after the stress begins to saturate, a finding
consistent with the experimental work of \cite{clementpng}.
The primary
contribution to stress fluctuations is randomness in the wall friction, a
boundary condition.
The fluctuations found in this model are set by a free parameter 
defining the magnitude of random friction. 
Our choice of this parameter results
in fluctuations of about $5\%$ of the mean stress, much less than
the $15 - 20\%$ found in experiments.

The results presented here are in qualitative agreement with
those of Socolar's $\alpha$-model.  His calculations incorporate
stress balance in both horizontal and vertical directions, and a balance of
angular momentum.
The essential feature of the
$\alpha$-model is that particle friction
transmits stress from particle to particle and, ultimately,
to the walls of a bin. 
These stresses, and any stress fluctuation, are partially
absorbed by the wall.  In contrast, the q-model
only considers vertical forces; stresses predicted by the q-model are
more like hydrostatic forces, and there is no mechanism for
dissipating fluctuations.

A difficulty faced by all of these models is correlations.
Experiments \cite{baxter} show chains of
particles experiencing high stress (the frequency of which
falls off exponentially with size).  These
pictures, and many other experiments,
suggest that grain forces are correlated.  
However, we lack adequate information to introduce correlations 
into models in a meaningful way.  
Experimental results reported by \cite{mueth}
measure static forces on short bins, and show no evidence for correlations.
The question of whether there are correlations, and over what
lengthscales are they important, is central to the entire formulation
of a continuum framework for granular materials.
Experimental and theoretical work is necessary to understand 
the nature of correlations.

Mueth et al.~\cite{mueth} also
study the frequency of fluctuations in three dimensional systems.
They find that, for fluctuations larger than the mean,
the frequency of fluctuation of a given size
decays exponentially with size of fluctuations.
For fluctuations smaller than the mean, the decay follows a power law.
Furthermore, their findings are largely unaffected by changes in the
boundary friction.  For purposes of comparison with this work, 
several factors are important.  The experimental set-up
has a depth-width aspect ratio of about 1 -- 1.5.
The glass beads and acrylic used in the experiment are very low
friction materials, with both internal and wall friction angles 
about $10-15^{\circ}$.  From the continuum perspective, 
stresses measured in this arrangement are hydrostatic-like.  Walls
do not support the bead pack, and even moderate 
changes in the wall friction
would have only minor effects on stress measurements.
We do not view these findings as invalidating the random friction
model proposed here, at least not for engineering applications.
The random packing model offers one possible
explanation of these experimental findings.
The mean stress for this model is given by Eqn.~3 in the limit
$\alpha \rightarrow 0$, and equals $\rho g y$; the
standard deviation, Eqn.~6 in the $\alpha \rightarrow 0$ limit,
is $\epsilon \rho g \sqrt{y}$.  For a short bin, $y \approx 1$,
and fluctuations are on the order of $\epsilon$ times the mean stress.
Packing variations, interpreted as voids fraction, can range up
to $20 - 30 \%$.  However even this model does not
explain all the physics of small aspect ratio bins.

\vspace{0.5in}

\noindent
{ACKNOWLEDGMENTS} I would like to thank Josh Socolar and Heinrich Jaeger
for providing me with preprints of their work, and Dave Schaeffer for
his comments about the manuscript.
\vspace{0.15in}

\eject

~

\begin{thebibliography}{99}

\bibitem{arnoldbook}
{\sc L. Arnold},
{Stochastic Differential Equations: Theory and Applications},
J. Wiley \& Sons, New York, 1974.

\bibitem{baxter}
{\sc G. W. Baxter},
{\it Stress Distributions in a Two Dimensional Granular Material}
in Powders and Grains '97, R. P. Behringer and J. T. Jenkins (eds.)
Balkema, Rotterdam, 1997, p. 345.

\bibitem{clementpng}
{\sc E. Clement, Y. Serero, J. Lanuza, J. Rajchenbach and J. Duran},
{\it Fluctuating Aspects of the Pressure in a Granular Column},
in Powders and Grains '97, R. P. Behringer and J. T. Jenkins (eds.)
Balkema, Rotterdam, 1997, p. 349.

\bibitem{sue1}
{\sc S.N. Coppersmith, C.-h. Liu, S. Majumdar, O. Narayan, and T.A. Witten},
{\it A Model for Force Fluctuations in Bead Packs},
Phys. Rev. E, {\bf53} (1996), p. 4673.

\bibitem{jansen}
{\sc H. A. Janssen},
{\it Versuche $\ddot u$ber Getreidedruck in Silozellen},
Zeitschrift. verein Deutscher Ingenieure {\bf 39} (1895), p. 1045.

\bibitem{sue2}
{\sc C.-h. Liu, S. R. Nagel, D. Shecter, and S. Coppersmith},
{\it Force Fluctuations in Bead Packs},
Science, {\bf 269} (1995), p. 513.

\bibitem{mueth}
{\sc D. Mueth, H. Jaeger and S. Nagel},
{\it Force Distribution in a Granular Material},
preprint.

\bibitem{neddermanbook}
{\sc R. M. Nedderman},
{Statics and Kinematics of Granular Materials}, Cambridge University Press,
Cambridge, UK 1992

\bibitem{schwedes}
{\sc J. Schwedes},
{\it Measurement of Flow Properties of Bulk Solids}
in Proc. of the International Symposium on Powder Technology 1981, 
p. 89.

\bibitem{socolar}
{\sc J. Socolar},
{\it Average Stresses and Force Fluctuations in Non-cohesive Granular
Materials},
preprint, available at xxx.lanl.gov as cond-mat/9710089

\bibitem{tadmor}
{\sc H. Nessyahu  and E. Tadmor},
{\it Non-oscillatory Central Differencing for Hyperbolic 
Conservation Laws},
J. Comp. Phys. {\bf 87} (1990), p. 408.


\end{thebibliography}
\end{document}